\documentclass[useAMS,usenatbib,epsf]{mn2e}
\begin{document}
\input{epsf}
\title[Fast H$_2$ in IRAS 16342-3814]{A fast bipolar H$_2$ outflow from IRAS~16342-3814: 
an old star reliving its youth}

\author[T.M. Gledhill et al.]
        {T.M.~Gledhill$^{1}$\thanks{email: {\tt t.gledhill@herts.ac.uk}},
         K.P.~Forde$^{1}$\\
         $^{1}$Science and Technology 
         Research Institute, University of Hertfordshire, 
         College Lane, Hatfield AL10 9AB, UK \\
         }
\maketitle

\begin{abstract}
Some evolved stars in the pre-planetary nebula phase produce 
highly-collimated molecular outflows that resemble the accretion-driven jets
and outflows from pre-main sequence stars. We show that
IRAS~16342-3814 (the Water Fountain Nebula) is such an object and
present $K$-band integral field spectroscopy revealing a fast ($>
150$~km~s$^{-1}$) bipolar H$_2$ outflow. The H$_2$ emission is
shock excited and may arise in fast-moving clumps, accelerated by the
previously observed precessing jet. The total luminosity in H$_2$ is
$0.37$~L$_{\odot}$ which is comparable with that of
accretion-powered outflows from Class~0 protostars. We also detect CO
overtone bandhead emission in the scattered continuum, indicating hot
molecular gas close to the centre, a feature also observed in a
number of protostars with active jets. It seems likely that the jet
and outflow in IRAS~16342-3814 are powered by accretion onto a binary
companion.
\end{abstract}


\section{Introduction}

Imaging studies of pre-planetary nebulae (pre-PN) at optical and
infrared wavelengths have revealed a wide range of bipolar, multipolar
and point-symmetric structures in scattered and thermal emission from dust 
(e.g. Lagadec et al. 2011; Si\'{o}dmiak
et al. 2008, Sahai et al. 2007; Gledhill 2005). These nebulae are
often accompanied by collimated outflows traced by the rotational
lines of CO and other molecules, and the near-infrared ro-vibrational
lines of H$_2$ (e.g.  Sahai et al. 2006; S\'{a}nchez Contreras et
al. 2004; Cox et al. 2003).  The process by which the outflows are shaped 
appears to involve, at
least in the case of the more extreme outflows, the appearance of
axial jets, which carve out cavities in the circumstellar material
(Sahai \& Trauger 1998).

This final flourish of activity by low- and intermediate-mass stars at
the end of the AGB is remarkably similar to the pre-main sequence
(pre-MS) outflow phase of their youth. Disc winds and jets are thought
to be an integral part of the star formation process and are seen from
the early Class~0 accretion phase through to Class~III and final
contraction onto the main sequence (e.g. Cabrit, Ferreira \& Dougados
et al. 2011). Pre-MS jets and outflows often manifest themselves as
shock-excited structures such as Herbig-Haro (HH) objects, visible in
CO and H$_2$ emission lines. Although the mechanism for ejecting and
driving pre-MS outflows is still a subject of debate, models fall
broadly into the categories of disc winds (K\"{o}nigl \& Pudritz 2000)
and X-winds (Shu et al. 2000), both of which involve
magnetocentrifugal outflows from accretion discs.  In the case of
pre-PN with bipolar outflows, it has been known for some time that the
momentum present in the outflow often greatly exceeds that which could
be supplied by a radiation-driven stellar wind (Bujarrabal et
al. 2001) so that an additional driving mechanism is required. It
seems likely that the additional energy may derive from the presence
of a binary companion, either stellar or substellar (review by De
Marco 2009), and that accretion may also play a central role in
powering these pre-PN outflows.

IRAS~16342-3814 (hereafter IRAS~16342), is one of the ``Water
Fountain'' sources which exhibit high-speed H$_2$O maser emission. The
masers in IRAS~16342 are thought to trace bow shocks at the ends of a
jet, with material in the head of the jet travelling at
155~km~s$^{-1}$ (Claussen, Sahai \& Morris 2009). Infrared imaging
shows a spectacular corkscrew feature which has been interpreted as
evidence of a precessing jet, actively carving cavities in the dusty
molecular envelope (Sahai et al. 2005).  In this paper we present
integral field spectroscopy in the $K$-band, using the SINFONI
instrument on the VLT, revealing a high-speed bipolar outflow in
H$_2$.  We draw comparisions with the well-studied pre-PN system
CRL~618 and then with pre-MS outflow sources which show strong
evidence for accretion-driven jets.

\section{Observations and data reduction}

Observations of IRAS 16342 were made on 2005 June 29 with the SINFONI
integral field spectrometer on the 8.2-m UT4 telescope at the VLT
observatory in Chile (Eisenhauer et al. 2003, Bonnet et al. 2004). The
{\it K}-band grating was used covering a wavelength range of $1.95$ to
$2.45~\umu$m with spectral resolution of $\sim 5000$. This equates to
a spectral pixel (channel) width of $2.45\times10^{-4}~\umu$m.  The
expected full width at half maximum (FWHM) in the spectral direction
is 2 channels or
$4.9\times10^{-4}~\umu$m ($\approx 66$~km~s$^{-1}$ at $2.2~\umu$m).

Medium Resolution Mode (MRM) was used providing a field of view of
3$\times$3 arcsec with $50\times100$~mas spatial pixels. The total
integration time on-source was 16~minutes. Offset sky exposures were
obtained along with the standard arc, flux and telluric
calibrations. Flux and telluric calibration were performed using
HD~156152. The average airmass was 1.23 with an ambient seeing of
$\approx 0.8$ arcsec. Adaptive optics (AO) correction in MRM resulted
in a slightly E-W elongated point spread function (PSF) of $0.13\times
0.10$~arcsec FWHM as measured using the standard.

Data reduction was accomplished using the ESO pipeline for SINFONI
with further processing and data visualization using the STARLINK
software collection. For further details on our processing of the
SINFONI data cubes, see Gledhill et al. (2011).

\section{Results and analysis}
\begin{table*}
\caption{Peak wavelength ($\lambda_{p}$), corresponding LSR velocity
($V_{\rm LSR}$)  and line flux ($F$) for the continuum-subtracted H$_2$ lines, shown
  separately for the NE and SW lobes. Flux units are
  $10^{-18}$~W~m$^{-2}$, wavelengths in microns and velocities in
  km~s$^{-1}$. The uncertainty on the velocity is
  estimated at 0.5 spectral channels, or $17~$km~s$^{-1}$.  The rest
  wavelength for each line ($\lambda_{0}$) is also given 
  (Bragg, Brault \& Smith 1982).}
\label{line-fluxes}
\begin{tabular}{lccccccc}
\hline 
Line    &$\lambda_{0}$ & \multicolumn{3}{c}{NE~lobe} & 
                                   \multicolumn{3}{c}{SW~lobe} \\
\hline
&  & $\lambda_{p}$ & $V_{\rm LSR}$ & $F$ & $\lambda_{p}$  & $V_{\rm LSR}$ & $F$ \\
\hline
1-0 S(3) & 1.9576 & 1.9585 &+139 &15.6 $\pm 1.0$ &1.9571 & -87&18.6 $\pm 1.2$  \\
1-0 S(2) & 2.0338 & 2.0350 &+174& 6.60 $\pm 0.81$ &2.0330 &-115& 9.37 $\pm 1.13$ \\
2-1 S(3) & 2.0735 & 2.0747 &+164& 1.71 $\pm 0.74$ &2.0730 &-84& 2.52 $\pm 1.06$ \\
1-0 S(1) & 2.1218 & 2.1229 &+150& 26.1 $\pm 0.9$ &2.1212 &-88&40.0 $\pm 1.5$ \\
1-0 S(0) & 2.2235 & 2.2246 &+144& 6.24  $\pm 0.76$ &2.2229 &-87&9.95 $\pm 1.31$ \\
2-1 S(1) & 2.2477 & 2.2491 &+183& 2.15  $\pm 1.06$ &2.2469 &-112&2.69 $\pm 1.89$ \\
1-0 Q(1) & 2.4066 & 2.4079 &+153& 28.8 $\pm 0.9$ &2.4059 &-91&43.0 $\pm 1.3$ \\
1-0 Q(2) & 2.4134 & 2.4147 &+160& 9.21  $\pm 0.91$ &2.4128 &-84& 11.7 $\pm 1.1$\\
1-0 Q(3) & 2.4237 & 2.4250 &+158& 27.3 $\pm 1.1$ &2.4231 &-85&39.7 $\pm 1.8$ \\
1-0 Q(4) & 2.4375 & 2.4387 &+147& 6.30  $\pm 0.97$ & 2.4368 &-94& 5.08 $\pm 1.35$ \\
\hline
\end{tabular}
\end{table*}

The pipeline-reduced and telluric-corrected datacube can be summed
over the 1.95 to 2.45~$\umu$m wavelength range of SINFONI to form a
``white light'' image, and this is shown in Fig.~1.  The
inner structure of the lobes is shown as a greyscale upon which we
superimpose contours to illustrate the fainter bipolar nebulosity. The
source is hidden behind the central region of obscuration which
separates the two lobes and we mark an approximate position as '+',
estimated as midway between the pinched isophotes in Fig.~1.  In fact
the source remains obscured even in mid-infrared observations with
$\tau \ge 100$ up to 30~$\mu$m (Verhoelst et al. 2009; Lagadec et
al. 2011).  In the $K$-band the bipolar lobes are seen primarily by
dust-scattered light, as indicated by their high degree of linear
polarization in this waveband (Murakawa \& Sahai 2010). 

Our image
bears strong similarity to the {\em K$_p$} ($1.948-2.299~\umu$m) Keck
image presented by Sahai et al. (2005) in which they identify
structure, including a number of knots, which they attribute to the
action of a precessing jet.
We indicate in Fig.~1 two knots in each lobe which are common between
our image and the $K_p$ image of Sahai et al. (2005), using their
nomenclature.  The integrated flux in the NE and SW lobes in our data
is $32.3\pm1.5$ and $75.2\pm1.5$~mJy respectively, which compares well
with the fluxes of 35 and 86~mJy measured by Sahai et al. (2005)
through their {\em K$_p$} filter. Our lower flux for the SW lobe is
easily explained by the fact that this lobe extends beyond the edge of
the SINFONI field of view and so is truncated in our observation
(Fig.~1).

\begin{figure}
\epsfxsize=8.4cm \epsfbox{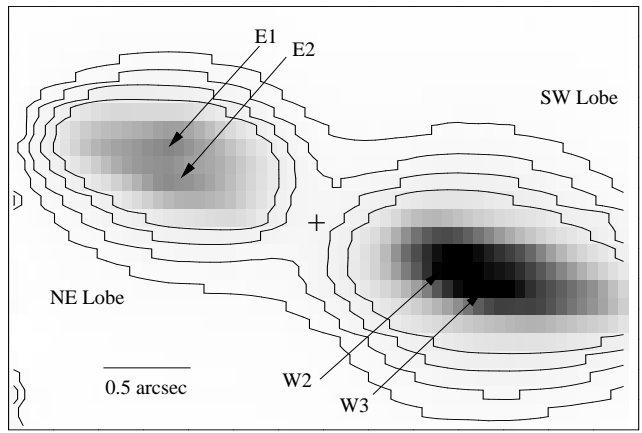}
\caption{A `white light' image of IRAS 16342 covering the
  wavelength range of $1.95-2.45~\umu$m and showing the inner
  structure in the lobes with contours of the fainter surrounding
  nebulosity.  The locations of the knots identified by Sahai et
  al. (2005) are indicated. Coutours are at 1.0, 1.6, 2.5, 4.0 with
  the peak of the greyscale at 55, in units of
  10$^{-15}$~W~m$^{-2}$~arcsec$^{-2}$. N is up and E to the left.}
\end{figure}

\subsection{$K$-band spectrum}
In Fig.~2 we show the spectrum of IRAS 16342 between 1.95 and
2.45~$\umu$m, showing detection of a number of H$_2$ ro-vibrational
lines, with fluxes listed in Table~1. A previous search for H$_2$
emission in 1993 by Garc\'{i}a-Hern\'{a}ndez et al. (2002)
resulted in a non-detection with a 2$\sigma$ upper limit of 
$5.1\times 10^{-17}$~W~m$^{-2}$. It seems likely that their wide
slit (4.5 arcsec) oriented at $-70$\degr~admitted too much continuum to detect the H$_2$
lines, although we cannot rule out the possibility that the H$_2$
emission has brightened between 1993 and 2005. The
continuum brightness in the SW lobe is approximately 2.5 times that in
the NE lobe, so we show the two spectra separately. Features present
in the scattered continuum are more evident in the SW lobe whereas the
H$_2$ emission lines are more evident in the NE lobe. In summary we note that: 
(i) the continuum rises steadily with
wavelength, indicating a very red source; 
(ii) the 1-0 S- and Q-branch
ro-vibrational transitions of H$_2$ are strong and we also detect a
number of 2-1 transitions. Upon
closer inspection, the H$_2$ lines are each doppler split into two
components, corresponding to blue-shifted (relative to the systemic
velocity) emission in the SW lobe and red-shifted emission in the NE
lobe. This is discussed further in Section~\ref{h2_emission}; 
(iii) a number of CO first overtone bandheads are seen in emission 
longward of 2.29~$\umu$m; 
(iv) there is a Na~I doublet feature in emission at 2.22~$\umu$m.

\begin{figure}
\epsfxsize=9.4cm \epsfbox[0 -25 395 466]{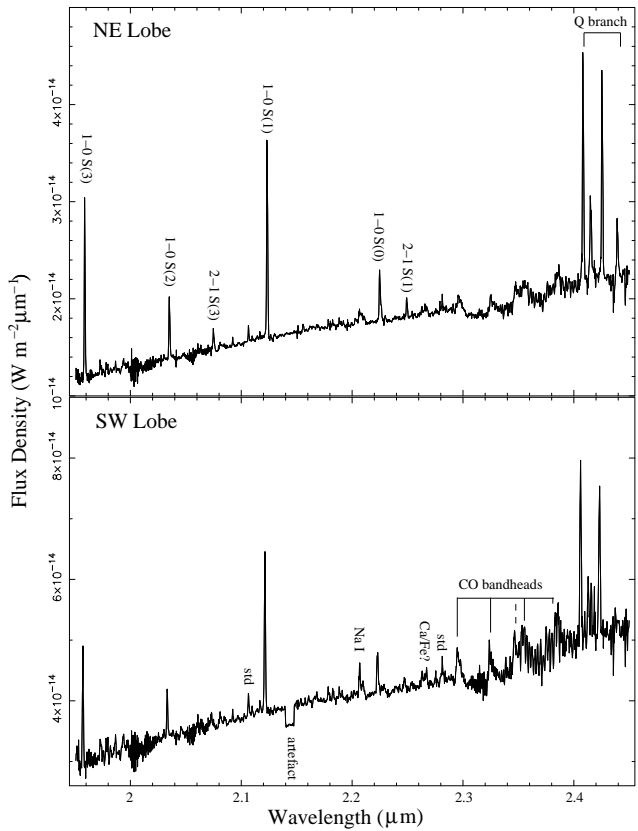}
\vspace{-10mm}
\caption{SINFONI spectra of IRAS 16342 integrated over the region of nebulosity
bounded by the third contour in Fig.~1. 
The continuum brightness differs by a factor of $\approx 2.5$ between the 
two lobes and so we show them separately. The H$_2$ emission lines are 
labelled in the top diagram and other features in the lower diagram. The
$^{13}$CO~$2-0$ bandhead is indicated with a dotted line. Features due to
the telluric standard are shown as ``std''.}
\end{figure}

\subsection{H$_2$ line ratios}
The $1-0$~S(1)/$2-1$~S(1) flux ratio is an important
excitation diagnostic and has values of $\sim 10$ and above for
shock-excited H$_2$ (e.g. Shull \& Hollenbach 1978). 
As the $2-1$~S(1) line is weak in IRAS 16342 (the flux in
this line is the same within errors in the two lobes) we
calculate a combined ratio for both lobes, giving 
$1-0$~S(1)/$2-1$~S(1)=$13.7\pm 6.1$, which is indicative of
shock-excited H$_2$. 

Any line of sight extinction will cause this ratio to increase,
remaining consistent with a shock interpretation.  An estimate of the
line of sight extinction from the H$_2$ emitting region to the
observer can be obtained from the ratio of the $1-0$~Q(3) and
$1-0$~S(1) lines and the assumption of a power law dependence of
extinction with wavelength across the $K$-band (e.g. Davis et
al. 2003). A value of $-1.75$ has often been assumed for the power law
exponent of the near-infrared extinction law, however we assume 
that
$A_{\lambda} \propto \lambda^{-2.14}$ for the $K$-band, which has been
estimated for the Galactic plane (Stead \& Hoare 2009) and is also
consistent with lines of sight towards the Galactic Centre
(Sch\"{o}del et al. 2010).  Using the $1-0$~Q(3) and $1-0$~S(1) line
strengths in Table~1 gives an extinction at the wavelength of the
$1-0$~S(1) line of $A_{\rm 1-0~S(1)}=1.77\pm 0.23$ and $1.54\pm 0.26$
magnitudes for the NE and SW lobes respectively. Although a higher
extinction towards the NE lobe would be consistent with the tilt of
the bipolar axis determined from previous studies (e.g. Claussen et
al. 2009) the two lobes have the same extinction within our
error. Using the combined fluxes for both lobes gives an extinction of
$A_{\rm 1-0~S(1)}=1.63\pm 0.18$ and an extinction-corrected line ratio
of $1-0$~S(1)/$2-1$~S(1)=$16.3\pm 9.1$.

A shock-excitation interpretation is also supported by the non-detection of
higher vibrational transitions, such as the $3-2$~S(3) line at
2.2014~$\umu$m.

\subsection{H$_2$ Excitation}
In Fig.~3 we plot a ro-vibrational diagram showing column
densities of H$_2$ for the lines in Table~1, relative to the
$1-0$~S(1) line. For a single temperature LTE gas the points will lie
along a straight line. We plot points for the NE and SW lobes
separately, each corrected with the extinction estimates derived in
the previous section. With the exception of the
$1-0$~Q(4) line in the SW lobe (at $T_{\rm upper}=7586$~K), the Q-branch and 
S-branch transitions deriving from the same upper energy level
agree well indicating that the correction for differential
extinction is good. The $1-0$~Q(4) flux appears anomalous (being the
only line in which the SW lobe is fainter than the NE lobe) and
suffers from a noisy continuum subtraction, lying at the
long-wavelength limit of the instrument.  A weighted least-squares fit
to the $v=1-0$ transitions, excluding this point,
gives a gas temperature of $1394\pm 51$~K. Although the $v=2-1$
transitions could be fitted within error by a line of similar slope,
they are clearly offset from the $v=1-0$ fit so that the vibrational
temperature is hotter by $400-600$~K (for S(3) and S(1) lines
respectively).
\begin{figure}
\epsfxsize=8.5cm \epsfbox[0 -160 543 630]{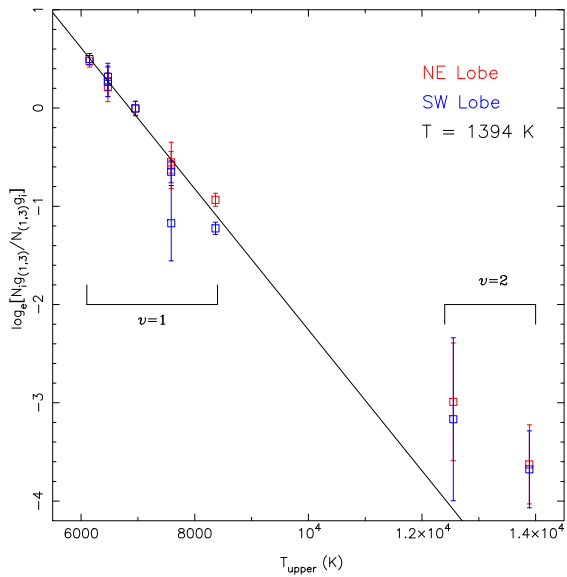}
\vspace{-4.7cm}
\caption{An excitation diagram showing the log of the normalized
  column densities against upper level temperature for the lines in
  Table~1. Red and blue symbols represent the NE and SW lobes
  respectively. A weighted least-squares fit to the $v=1-0$
  transitions (excluding the $1-0$~Q(4) SW lobe point) 
corresponds to a LTE gas temperature of 1394~K.}
\end{figure}
An offset between the vibrational levels can result in regions where
the shocked gas is not fully thermalised and this is seen in 3D models
of bow shocks (e.g. Gustafsson et al.  2010). There is also evidence for
a similar offset in the H$_2$ ro-vibrational diagram for CRL~618, a
pre-PN also with a fast H$_2$ outflow (Thronson 1981) and discussed further in 
Sec.~4.2.  

\subsection{H$_2$ spatial distribution}
\label{h2_emission}
\begin{figure}
\epsfxsize=8.5cm \epsfbox{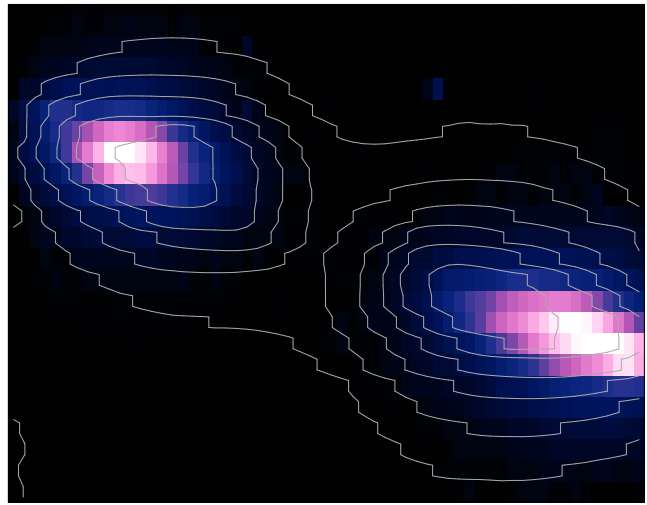}
\caption{An image of the $1-0$~S(1) line emission with 
  $K$-band white light contours (from Fig.~1), showing the relative
  locations of the H$_2$ and the scattered continuum. The peak of the
  colour scale (white) corresponds to $1.9\times
  10^{-16}$~W~m$^{-2}$~arcsec$^{-2}$.}
\end{figure}
The relative locations of the H$_2$ emission and the scattered
continuum are shown in Figs.~4 and 5 for the $1-0$~S(1) line (other
lines show a similar structure). In the accompanying spectrum to
Fig.~5 the line is clearly split into blue- and red-shifted
components, relative to a central wavelength of 2.1222~$\umu$m. A
similar split is seen in the other H$_2$ lines. The blue-shifted
emission is associated with the SW lobe and the red-shifted emission
with the NE lobe (a small amount of blue (red) emission appears in the
NW (SE) lobe close to the central wavelength due to the spectral PSF
of SINFONI). The peak of the H$_2$ emission in each lobe is located at
a greater angular offset from the centre than the knots, which are
thought to represent enhanced scattering from denser material swept up
in the lobe walls. The peaks occur close to the axis of the system at
PA 68\degr, defined by the scattered light image, but their offset
along this axis is not symmetrical with respect to the centre of the
obscuring lane (marked '+'); the NE (red shifted) H$_2$ peak lies at
an offset of 1.0 arcsec whereas the SW (blue shifted) peak forms an
elongated feature lying between 1.3 and 1.5 arcsec offset. However the
continnum contours show that the lobes are not of equal size, with the
larger SW lobe extending beyond our field of view. Relative to the
outer continuum contour, the H$_2$ peaks do seem to occupy a
symmetrical position, being set back by approximately 0.56 arcsec from
the tips of both lobes (estimated in the case of the SW lobe). It is
notable from the contours in Fig.~5 that the peak H$_2$ surface
brightness is very similar in both lobes, despite the continuum
brightness being a factor 2.5 higher in the SW lobe (contours in Fig.~4).

\begin{figure}
\label{fig_h2}
\epsfxsize=9cm \epsfbox{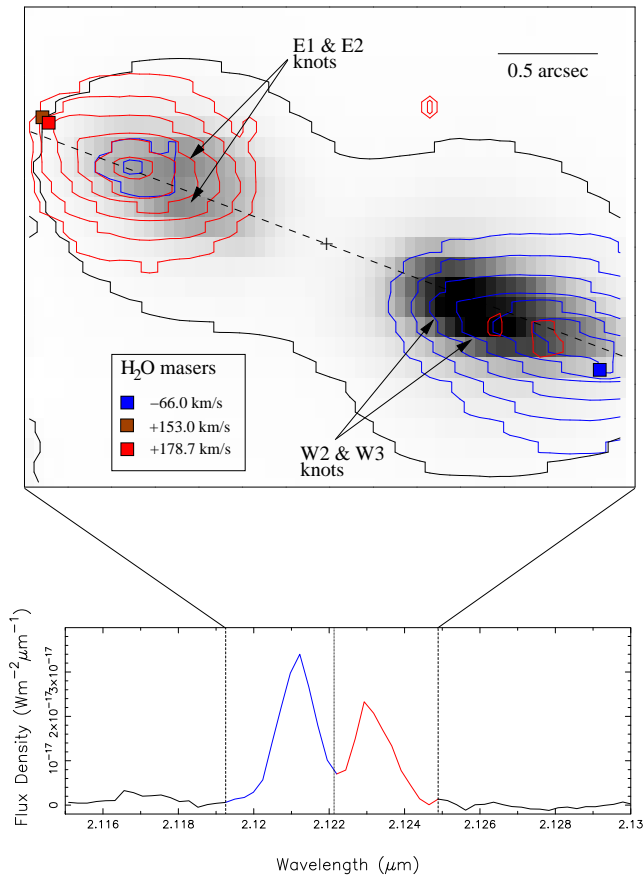}
\vspace{-3cm}
\caption{Contours of 1-0~S(1) H$_2$ emission, extracted and colour
  coded according to the spectrum in the lower panel, superimposed on
  the greyscale white-light image from Fig.~1. The outer contour from
  Fig.~1 is included to show the faint scattered light structure, with
  the symmetry axis shown as a dotted line at PA 68\degr. The centre
  of the $K$-band obscuration is marked '+'.  The systemic velocity of
  42~km~s$^{-1}$ derived from maser observations is shown as a dotted
  line on the spectrum. H$_2$ contour levels are 2, 4, 8, 16, 32,
  63 $\times 10^{-17}$W~m$^{-2}$~arcsec$^{-2}$. The locations of the
  VLBA H$_2$O masers from Claussen et al. (2009) are shown.}
\end{figure}

The locations of the $V_{\rm LSR}$=+153,
$+178.7$ and $-66$~km~s$^{-1}$ H$_2$O maser clumps
detected in 2002 VLBA observations (Claussen et al. 2009) are shown in
Fig.~5. Using the proper motions established by these multi-epoch
measurements, we have adjusted the maser positions to account for the
3 year interval between the VLBA and SINFONI observations. 
We set the origin in our data as the centre of
the obscuring lane and plot the maser positions relative to this point
(marked `+' in Fig.~5). All three H$_2$O maser clumps are located
close to the axis of the nebula, as indicated by the dotted line, but
are slightly offset to the N (S) in the NE (SW) lobes in a
point-symmetric fashion, and lie along an axis at PA 66\degr.  These
offsets are interpreted by Claussen et al. (2009) as further evidence
for the presence of a precessing jet.

\subsection{H$_2$ kinematics}
The spectral separation between the two $1-0$~S(1) line peaks (7
SINFONI channels, or $1.715\times 10^{-3}~\umu$m) corresponds to a
velocity separation of 238~km~s$^{-1}$. The LSR velocities of the line
peaks are $-88$ and $+150$~km~s$^{-1}$ with a central velocity (midway
between the two peaks) of $+31$~km~s$^{-1}$. A cautious error estimate
of half a spectral pixel corresponds to $\pm 17$~km~s$^{-1}$.
Velocities of the various line peaks
are given in Table~1.

A systemic velocity for IRAS~16342 of $43.2\pm 0.9$~km~s$^{-1}$ was
determined by Likkel, Morris \& Maddalena (1992), by searching for
velocity-symmetric H$_2$O maser pairs, in agreement with Likkel \&
Morris (1988) and te Lintel Hekkert et al. (1988). Sahai et al. (1999)
estimate $42\pm 2$~km~s$^{-1}$ from the OH masers, and the more recent
H$_2$O maser study by Claussen et al. (2009) assumes this value.
Taking the systemic velocity and outflow axis inclination as $+42\pm
2$~km~s$^{-1}$ and 45\degr~ from the maser observations, then the
H$_2$ outflow velocity relative to the source (i.e. along the bipolar
axis) is $153$ and $184\pm 24$~km~s$^{-1}$ in the NE and SW lobes
repectively. A higher outflow velocity in the SW lobe may be
consistent with the H$_2$-emitting region in that lobe being located
further ($1.3-1.5$~arcsec) from the source than in the NE lobe
($1.0$~arcsec), although within our errors the outflow velocities in
the two lobes could be the same.

The $+153$~km~s$^{-1}$ maser clump at the tip of the NE lobe shares a
very similar velocity with the red-shifted peaks of the bright
$1-0$~S(1), Q(1) and Q(3) H$_2$ lines ( $V_{\rm LSR}=+150$, $+153$ and
$+158$~km~s$^{-1}$ respectively). The other red-shifted H$_2$ lines
also have velocities consistent with $+153$~km~s$^{-1}$ to within $\pm
17$~km~s$^{-1}$ ($\pm 0.5$ spectral channel), apart from the
$2-1$~S(1) line, which has a peak at $+183$~km~s$^{-1}$, and the
$1-0$~S(2) line at $+174$~km~s$^{-1}$. The latter two lines are more
comparable with the $+178.7$~km~s$^{-1}$ maser (see Table~1).  These
velocity associations in the NE lobe suggest that the H$_2$O maser and
H$_2$ emission may originate within gas sharing similar bulk
motion. In the SW lobe, the bright $1-0$~S(1), Q(1) and Q(3) lines
peaks are blue-shifted at $V_{\rm LSR}=-88$, $-91$ and
$-85$~km~s$^{-1}$ respectively.  Although we have not imaged the tip
of this lobe, we appear to have imaged the peak of the H$_2$ emission, which
lies close to the blue-shifted maser clump at $-66$~km~s$^{-1}$, again
suggesting similar bulk motion in the gas responsible for the H$_2$
and H$_2$O emission. As in the NE lobe, the $1-0$~S(2) and $2-1$~S(1)
lines show higher velocities.

\subsubsection{Line profiles}
\label{lineshape}

To investigate the H$_2$ kinematics further we plot a
position-velocity (P-V) diagram obtained by simulating an 8-pixel
(0.4~arcsec) wide slit oriented along the major axis of the nebula
(the dashed line in Fig.~5). This is shown in Fig.~6 (centre) along
with a number of velocity slices taken at various offsets along the
`slit' (top and bottom). We note the following features:
\begin{enumerate}
\item The slices show that the peak flux density in the $1-0$~S(1)
  line is approximately the same in both lobes (panels b and e)
  despite the continuum peak in the SW lobe being a factor 2.5 times
  brighter than in the NE lobe. This suggests that the temperature and
  column density of shocked H$_2$ are similar in both lobes and that
  the higher continuum flux in the SW lobe is due to a
  forward-scattering dust phase function rather than an increase in
  dust and gas density.
\item The line broadens as we move from the tips of the lobes
  towards the source (from $a\rightarrow c$ and from $f\rightarrow d$). The
  line profiles in Fig.~6 indicate a wide range of velocities, with
  FWHM $\sim 200$~km~s$^{-1}$.
\item The line becomes increasingly asymmetric moving from the lobe
  tips towards the source, so that in the NE (red shifted) lobe it
  develops a red wing (panels $a\rightarrow c$) whereas in the SW
  (blue shifted) lobe a blue wing develops (panels $f\rightarrow d$),
  eventually splitting into two velocity components in the P-V
  diagram. 
\item We do not see any clear evidence for a shift in the line peak 
(increase in outflow velocity)
  with offset from the star.
\end{enumerate}

\begin{figure}
\label{pv_h2}
\epsfxsize=9cm \epsfbox{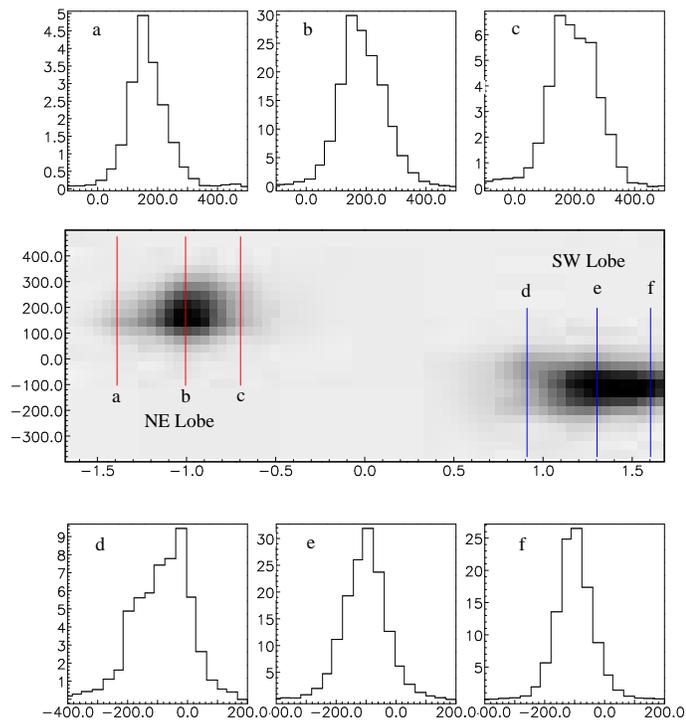}
\caption{The central panel shows a P-V diagram with arcsec offset along the
the horizontal axis and $V_{\rm LSR}$ in km~s$^{-1}$ along the vertical axis. 
Slices of H$_2$ surface brightness in the 
velocity direction (km~s$^{-1}$) are shown at six positions along the outflow axis. 
The H$_2$ surface brightness is in units
of $2\times 10^{-19}$~W~m$^{-2}$~$\umu$m$^{-1}$~arcsec$^{-2}$.}
\end{figure}

Schultz et al. (2005) present a gallery of H$_2$ line profiles for bow
shocks with a range of geometries, inclinations and shock velocities.
These show line widths $\sim V_{\rm bow}$, where $V_{\rm bow}$ is the
velocity of the bow relative to the ambient medium, so that fast bow
shocks are expected to produce broad H$_2$ emission lines.  In the
case of an inclined outflow axis (45\degr~for IRAS~16342, Claussen et
al. 2009), then the line profile can become asymmetric with a
high-velocity wing arising from the bow apex and the core of the line
resulting from the larger area of lower-velocity gas in the near
surface of the shock. Although asymmetric line profiles are evident in
our data (Fig.~6), the high velocity wings become stronger at
positions $c$ and $d$, {\em further} from the bow apex and closer to
the source, which is hard to interpret as a simple terminal bow
shock. This is discussed further in Sec.~4.3.

\subsection{Mass of the H$_{2}$ outflow}
The flux in the $1-0$~S(1) line can be used to estimate the mass of
shocked H$_2$ in the outflow, following the procedure outlined by
Davis et al. (2001). For a gas excitation temperature of 1400~K (Sec.~3.2),
the column density of H$_2$ in m$^{-2}$ is given by $N_{\rm H_2}=7.596
\times 10^{28} I_{\rm 1-0 S(1)} 4/(a\pi)$, where $I_{\rm 1-0 S(1)}$ is
the extinction-corrected flux in the $1-0$~S(1) line in W~m$^{-2}$ and
$a$ is the solid angle subtended by the H$_2$-emitting region at the
observer. Multiplying by the area of the H$_2$ emitting region (in
m$^{2}$) and the mass of the H$_2$ molecule then gives the mass of
shock-excited H$_2$, 
which can be expressed as $M_{\rm
  H_2}=1.620\times 10^{5} I_{\rm 1-0 S(1)}d^{2}$~M$_{\odot}$ where $d$
is the distance to the object in pc. Taking the flux in both lobes
(Table~1) and correcting for an extinction at $2.122$~$\umu$m of $1.63
\pm 0.3$~mag. (Sec.~3.1) gives $I_{\rm 1-0 S(1)}=2.97 \pm 0.83 \times
10^{-16}$~W~m$^{-2}$.  
The distance estimate for IRAS~16342 ranges from
700~pc (Zijlstra et al. 2001) to 2~kpc (Sahai et al. 1999). The latter
was assumed by Claussen et al. (2009) to determine the tangential
H$_2$O maser velocities from their proper motions, which led to an
estimate of 45\degr~for the outflow axis, in agreement with
independently-derived values. A distance of 700~pc would lead to a
much higher inclination (70\degr). The total mass in 
H$_2$ is then $M_{\rm H_2}=1.92 \pm 0.54
\times 10^{-4}(d/{\rm 2 kpc})^2$~M$_{\odot}$, with the relatively large
error arising from the uncertainty on the extinction and distance
estimates. The linear extent of the H$_2$-emitting lobes is $\approx
1.3$ arcsec which, correcting for the inclination of 45\degr, will be
traversed in $\approx 112$~years by an outflow travelling at
$155$~km~s$^{-1}$, giving a mass flux of $\dot{M}_{\rm H_2}=1.7 \pm 0.7 \times
10^{-6}(d/{\rm 2 kpc})$~M$_{\odot}$~yr$^{-1}$ in H$_2$, assuming an
uncertainty of 20~km~s$^{-1}$ on the outflow velocity. For a gas
with temperature $T=2000$~K then $M_{\rm H_2}$ and
 $\dot{M}_{\rm H_2}$ are a factor $\sim 3$ lower. 

\subsection{CO bandhead and Na~{\sevensize{I}} emission}
We detect the $v=2-0, 3-1, 4-2$, and $5-3$ $^{12}$CO overtone
bandheads in emission. We also see the $v=2-0$~$^{13}$CO bandhead,
again in emission as indicated in Fig.~2.  The presence of the
$^{13}$CO bandhead is consistent with the low $^{12}$CO/$^{13}$CO line
ratio from millimetre data obtained by He et al. (2008) and
interpreted as possible evidence for hot-bottom burning and hence a
high-mass progenitor for IRAS~16342. We estimate that the flux in the
$v=2-0$ and $3-1$~$^{12}$CO features is $1.4\pm 0.3$ and $2.6\pm
0.5 \times 10^{-17}$~W~m$^{-2}$ with large uncertainties due to the
difficulty in subtracting a continuum in this region.

A Na~{\sevensize{I}} doublet feature is seen in emission with peaks at
2.2069 and 2.2098~$\umu$m. The integrated flux across the feature is
$1.24 \pm 0.16 \times 10^{-17}$~W~m$^{-2}$.

Our integral field observations show that these spectral features have
the same spatial distribution as the continuum and hence originate
from a compact source in the central region, obscured from direct
view, and are scattered into our line of sight by dust in the bipolar
lobes. This is in contrast to the H$_2$ emission which traces the
extended bipolar outflow. The situation is remarkably similar to that
of IRAS~18276-1431 discussed by Gledhill et al. (2011), where the CO
and Na~{\sevensize{I}} emission was interpreted as evidence for the
presence of hot ($> 2000$~K) and dense ($> 10^{10}$~cm$^{-3}$) gas close 
to the central star(s).

\section{Discussion}

\subsection{CO rotational emission in IRAS 16342}
CO emission has been detected toward IRAS~16342 in both the $J=2-1$
and $J=3-2$ lines (He et al. 2008; Imai et al. 2009). Although
spatially-resolved observations are not yet available, association
with interferometric maser data on the basis of velocity gives an
indication of the likely CO structure. He et al. (2009) associate the
$J=2-1$ emission (expansion velocity 46~km~s$^{-1}$) with the OH main-line
masers which trace the base of the optical/infrared bipolar lobes
(Sahai et al. 1999).  The larger CO $J=3-2$ velocity extent of
158~km~s$^{-1}$ (Imai et al. 2009) is more comparable with that of the
OH 1612~MHz masers (130~km~s$^{-1}$; Sahai et al. 1999; Zijlstra et
al.  2001) which are located further from the star, along the edges of
the NE and SW lobe. This then gives the picture of CO gas accelerating 
with offset from the star. 

Imai et al. (2009) derive a mass-loss rate of $\dot{M}_{\rm
  gas}=2.9\times 10^{-5}$~M$_{\odot}$~yr$^{-1}$ from the CO $J=3-2$
line. This compares with $1.7\times 10^{-5}$~M$_{\odot}$~yr$^{-1}$
obtained by He et al. (2008)\footnote{after a correction applied by
  Imai et al.  (2009)} from CO $J=2-1$ measurements. Both values are
greater, by at least a factor of 10, than our estimate from the H$_2$
emission. However, as noted by Imai et al. (2009) the CO mass-loss
rates are almost certainly over-estimated due to the assumption of
spherical symmetry.  Conversely, it is likely that our value of
$\dot{M}_{\rm H_2}=1.7\times 10^{-5}$M$_{\odot}$~yr$^{-1}$ (Sec.~3.4)
will be an underestimate since it only accounts for the component of
cooling H$_2$ that has been shock heated by the jet.

Imai et al. (2009) devlop a spatio-kinematic (SHAPE) model to fit the
CO $J=3-2$ line profile and conclude that, while their model is not
unique, the emission is consistent with a bipolar CO flow with outflow
velocity proportional to distance from the star and an exponentially
decreasing gas density.  The CO and OH emission then traces material
that is being accelerated and swept-up into the bipolar cavity walls
by the high-speed axial jet, transferring momentum as it goes.

In contrast, the H$_2$ emission shows no evidence for a low-velocity
component and instead appears to be associated purely with the high
velocity axial jet, with velocities in both lobes comparable to the
H$_2$O masers at the jet tips. Also, there is no clear evidence in
H$_2$ for the velocity-offset relation seen in the CO and OH data,
which is again consistent with the H$_2$ emission arising in an
axial jet rather than in accelerating swept-up cavity wall
material.

\subsection{Comparison with CRL 618}
\subsubsection{H$_{2}$ emission}
The spatial and kinematic structure of the H$_2$ emission in
IRAS~16342 bears a number of resemblances to that of the bipolar
proto-PN CRL~618, which also displays shock excited H$_2$ 
(Thronson 1981)\footnote{although given
  the B0 or earlier central star and centrally located H{\sevensize
    II} region in CRL~618 (Westbrook et al. 1975; Schmidt \& Cohen
  1981) a fluorescent component in this object also seems likely}.
Firstly, Cox et al. (2003) map the $1-0$~S(1) line in CRL~618 with 0.5
arcsec and 9~km~s$^{-1}$ resolution and find that the
velocity-integrated emission has a lobe-filled structure, as seen in
IRAS~16342, with the emission peaks located close to the nebula axis,
rather than along the edges of the lobes.
Secondly, high velocity wings in the $1-0$~S(1) line were noted in CRL~618 by
Burton \& Geballe (1986) with a line width of $\approx
250$~km~s$^{-1}$ (c.f. $\approx 240$~km~s$^{-1}$ line peak separation
in IRAS~16342; Sec 3.2.1), indicating a high-velocity outflow.
Spectral imaging with a velocity resolution of 4~km~s$^{-1}$ (Kastner
et al. 2001) shows that H$_2$ emission up to 120~km~s$^{-1}$ (line of
sight) is present in both lobes. This is confirmed by Cox et
al. (2003) who show that the high-velocity H$_2$ is associated with
jets traced in optical data, with emission peaking behind the jet
heads and extending downstream towards the star. Velocities of
140~km~s$^{-1}$ (relative to systemic) behind the jet head in the E
lobe correspond to an outflow velocity of 260~km~s$^{-1}$ assuming an
inclination of 32\degr~relative to the plane of the sky (S\'{a}nchez
Contreras et al. 2004; hereafter SC04). This jet speed is higher than
that of IRAS~16342, where the average of the two lobes is
169~km~s$^{-1}$ from the H$_2$ data (Sec.~3.2.1) and 153~km~s$^{-1}$
from the H$_2$O masers (Claussen et al. 2009).

There are also some notable differences between the two objects.
The H$_2$ position-velocity (P-V) diagram for CRL~618 (Cox et al.
2003) shows a characteristic ``butterfly'' structure, resulting from
the presence of both an extended low-velocity component and a high
velocity component which terminates closer to the star (within $\pm
2$~arcsec along the axis). Although our velocity resolution is much
lower, we are confident of the absence of such a butterfly structure 
in the P-V diagram of
IRAS~16342. The line widths do broaden closer to the star, but
there is no evidence for a low-velocity component. In CRL~618,
low-velocity H$_2$ emission is associated with the expanding cavity
walls and high-velocity emission with the optical jets.  In IRAS~16342
the H$_2$ emission appears to be exclusively associated with the
high-velocity axial jet.

\subsubsection{CO emission}
The characteristic butterfly P-V diagram structure, seen in H$_2$
emission in CRL~618, is also seen in interferometric observations of
CO~$J=2-1$ (SC04) and to a lesser extent CO~$J=6-5$ (Nakashima et
al. 2007). The CO~$J=2-1$ maps show two outflow components: a fast
axial bipolar outflow, peaking within 2 arcsec of the star, and a more
spatially extended lower velocity flow which appears limb brightened
and traces the optical lobes. SC04 reproduce the CO~$J=2-1$ P-V diagram
and spatial distribution with a model consisting of a fast (up to
270~km~s$^{-1}$) cylindrical molecular jet which blows along the axis
of more slowly expanding (22~km~s$^{-1}$) bipolar elliptical
cavities. The high-velocity component (corresponding to the extended
wings of the P-V diagram) is created by the axial jet whereas the lower
velocity and more spatially extended emission arises from shocks at
the surfaces of the expanding cavities.  In the case of IRAS~16342, it
is not yet clear whether any of the high-velocity CO emission is
associated with the jet.

\subsection{Jet models}
Lee, Hsu \& Sahai (2009), building on earlier models of the
optical emission (Lee \& Sahai 2003), attempt to model the CO and
H$_2$ emission in CRL~618 using both atomic and molecular winds
interacting with a spherically symmetric AGB envelope. They find that
(i) it is difficult to produce the observed high-velocity CO and H$_2$
emission along the axis unless the fast wind is intrinsically
molecular and (ii) the wind must be collimated, possibly cylindrical
as assumed by SC04 (above),
rather than radially expanding as in their model. The
presence in IRAS~16342 of both fast axial H$_2$ emission (this paper),
and slower CO emission consistent with expanding cavity walls (Imai et
al.  2009), suggests a similar model in which an axial jet sweeps up
and accelerates material into the walls of bipolar cavities. Such a
model is also consistent with the OH maser data (Zijlstra, Matsuura \&
Dijkstra 2004).  The requirement for collimation raises the
possibility that the jet may be magnetized (MHD jet models in pre-PN are
discussed by Garc\'{i}a-Segura, Lopez \& Franco 2005 and references
therein). Highly-polarized OH maser emission has been detected in the
blue-shifted part of the 1612~MHz spectrum (Szymczak \& G\'{e}rard
2004), implying the presence of magnetic fields in the SW lobe at least.

\subsubsection{Precessing jets}
The inscribed corkscrew
pattern seen in NIR images (Sahai et al. 2005) provides evidence for the
presence of a precessing jet in IRAS 16342, with at least 1.5
rotations visible in the NE lobe (knots E4 to E1) and 3 in the SW lobe
(knots W1 to W4).  This structure, seen in scattered light, is thought
to result from the jet acting as a snow plough, piling up dusty
material into a helical density enhancement, as it expands away from
the source. Using an average offset between adjacent turns in the
corkscrew feature of 0.28~arcsec, $d=2$~kpc, an outflow speed of 90
km~s$^{-1}$ and an
inclination of the symmetry axis to the line of sight of $i=40\degr$,
Sahai et al. (2005) estimate that the precession period is $t_{p}\le
50$~yr. Taking instead our H$_2$ outflow velocity of 155~km~s$^{-1}$
as typical of the jet propagation speed and $i=45\degr$ from the
H$_2$O maser observations, then $t_{p}=24$~yr. This faster precession
speed is more compatible with the number of observed turns in the jet
and a jet dynamical age of $110-130$~yr inferred from the H$_2$O
masers (Claussen et al. 2009).


Hydrodynamic models of molecular jets have been produced to simulate
observed molecular outflows from protostellar sources (e.g. Suttner et
al. 1997; Smith, Suttner \& Yorke 1997). Models with both slow
($t_{p}=400$~yr) and fast ($t_{p}=50$~yr) precessing jets are
described by Smith \& Rosen (2005) and Rosen \& Smith (2004)
respectively. These models are able to reproduce some of the basic
features of the molecular observations of IRAS~16342 and CRL~618, such
as the presence of $1-0$~S(1) emission within the body of the
outflow surrounded by lower excitation CO~$J=2-1$ delineating the
outflow lobes. The fast-precessing jet model of Rosen \& Smith (2004)
seems most closely matched with IRAS~16342, having a jet velocity of
100~km~s$^{-1}$, precession period of $t_{p}=50$~yr
and precession angles (half angle of the precession cone) of 5, 10 and
$20\degr$.  Using the locations of the near-infrared knots we estimate the
precession angle for IRAS~16342 to be $\approx 10$\degr~when corrected
for the system inclination of 45\degr~to the plane of the sky.  The
jet diameter is assumed to be 114~au in the models, which is very similar 
to the $\le 100$~au deduced by Sahai et al. (2005) from the width of the corkscrew
pattern. Rosen \& Smith present synthetic images of H$_2$ and CO
emission which show that the $1-0$~S(1) emission develops as an
annular structure before breaking into multiple bow shocks. Low
excitation CO (e.g. $J=2-1$) traces the walls of the
precessional cone with higher excitation lines (e.g.  $J=6-5$)
contributing to emission from the jet. This is similar to the
structure observed in CRL~618 (see fig.~1 of Lee et al. 2009).  In the
case of IRAS~16342, resolved imaging of the CO lines is needed to
determine whether its molecular outflow is similar to that of CRL~618,
although the model of Imai et al. 2009 would be consistent with
this picture.

In the Rosen \& Smith (2004) models, the action of a narrow precessing
jet is to excavate an annular region rather than a hollow cavity,
leaving a core of relatively undisturbed material along the axis. This is at odds
with observations which suggest that the lobes in IRAS~16342 have
interiors that are tenuous compared with their walls (Sahai et
al. 2005). This may indicate that the cavities in IRAS~16342 were not
in fact created by the jet, but predate it.

\subsubsection{H$_2$ dissociation and fast-moving clumps}
A further problem with the shock models as they stand is that they do not
produce the high-velocity H$_2$ emission observed in IRAS~16342,
CRL~618 and many YSO outflows, due to the dissociation of H$_2$
molecules at hydrodynamic shock speeds of $\le 24$~km~s$^{-1}$ in
dense media (e.g. Hollenback \& McKee 1980). The problem has been
discussed widely in connection with YSO outflows and was highlighted
in the case of CRL~618 by Burton \& Geballe (1986) who advanced four
possibilities for the very broad S(1) line profile in that object: (i)
upscattering of H$_2$ photons off fast-moving dust grains; (ii) MHD
shocks; (iii) reformation of H$_2$ on dust grains behind a fast-moving
shock; (iv) H$_2$ emission from fast-moving clumps or bullets.

We discount the first possibility as in order to observe the clearly
defined red- and blue-shifted outflows then the dust would have to be
moving away from the H$_2$-emitting region in the NE lobe and toward
it in the SW lobe, which seems contrived.  In addition, the H$_2$
emission and scattered continuum peaks are spatially separated.  MHD
shocks are likely in an object with a highly-collimated jet, however,
the critical velocity is only raised to $\simeq 50$km~s$^{-1}$
(e.g. Smith 1994) which is still far too low to explain the
observations.  Reformation of H$_2$ behind a fast, dissociative shock
can occur on a timescale of $10^{16}/(n~{\rm cm}^{-3})$~s (Smith,
Khanzadyan \& Davis 2003), or within a dynamical timescale of $\sim
100$~yr for IRAS~16342 if $n>3\times 10^{6}$~cm$^{-3}$. The reformed
H$_2$ would then have to be heated to $1400$~K in subsequent
shocks in order to produce the observed line ratios.

Burton \& Geballe (1986) favoured the fast-moving clump or ``bullet''
model for CRL~618, and this was also supported by Ueta, Fong \&
Meixner (2001) based on the clumpy nature of the NIR images. It is not
clear yet whether fast moving clumps are responsible for the
high-velocity H$_2$ emission in IRAS~16342 but the model has
attractions. Firstly, a clumpy structure for the outflow is consistent
with the high-velocity H$_2$O masers, which require local densities
$\sim 10^{9}$~cm$^{-3}$ (Elitzur 1992). The fact that the H$_2$ emission 
has similar kinematics to the masers (Sec.~3.4) implies that it too
may arise in these clumps.
Secondly, a clumpy outflow may
offer an explanation for the asymmetric line profiles mentioned in
Sec.~3.4.1, where a blue wing develops upstream of the H$_2$ peak in
the blue-shifted lobe and a red wing in the red-shifted lobe: clumps of
material, accelerated by the jet, will move ballistically,
i.e. radially outward, in a direction determined by the jet
orientation at that particular point in its precessional cycle when
the interaction between the jet and the clump occurs. In Fig.~5 the
contours of H$_2$ emission curve up (down) at the ends of the NE (SW)
lobes, towards the H$_2$O maser positions, which lie off the nebula
axis in a point-symmetric fashion.  This presumably indicates the
current projected orientation of the jet.  Moving back along the
nebula axis in the NE lobe, and hence back along the precessional
spiral, we can speculate that the jet axis points predominantly behind
the plane of the sky here, so that a red wing develops on the line
(moving from $a$ to $c$ in Fig.~6) as the line of sight includes
clumps with higher radial velocity. If the jet is point symmetric,
then at an equivalent position in the SW lobe, its axis would be
pointing towards the front surface of the lobe, producing a
blue-shifted wing to the line, which would diminish towards the
end of the blue lobe. However, the asymmetries in the line profiles
appear quite pronounced, extending over several SINFONI spectral
pixels in Fig.~6, whereas the shift in radial velocity corresponding
to a precession angle of 10\degr~amounts to only $\sim 1$ spectral
channel for
a jet speed of 155~km~s$^{-1}$. An additional source of asymmetry
appears necessary and further observations with higher velocity- and
spatial-resolution are needed to investigate these possibilities
further.

\subsection{Comparison with pre-MS outflows}
Collimated outflows are thought to be an integral part of star
formation, extracting angular momentum from a circumstellar disc and
enabling accretion onto the protostar. In Class~O YSOs (age $\le
10^{4}$~yr) the outflows are mainly molecular, often with a fast
($10-100$~km~s$^{-1}$) and highly collimated jet, traced in H$_2$,
driving a slower ($\sim 10$~km~s$^{-1}$) and wider-angle outflow of
swept-up gas, traced by CO (see Cabrit, Ferreira \& Dougados 2011 for
a recent review). There are clearly similarities with post-AGB
outflows such as IRAS~16342, CRL~618 and IRAS~22036+5306 (Sahai et
al. 2006). The Class 0 outflow of HH~211 can be compared to IRAS~16342
including (i) a deeply-embedded driving source detected only at
sub-millimetre wavelengths, (ii) outflow cavity structure traced in
low-velocity CO emission with a high velocity axial component (Gueth
\& Guilloteau 1999) and (iii) shocked H$_2$ emission consistent with
interaction of a jet with the ambient material (McCaughrean, Rayner \&
Zinnecker 1994). The HH211 jet itself is mapped in SiO~$J=8-7$ emission as a
spectacular chain of knots extending $>15$~arcsec on either side of
the source into the H$_2$-emitting region (Lee et al. 2009b).  Wiggles
in the jet structure are interpreted as orbital motion of the jet
source in a binary system, rather than precession (Lee et al. 2010)
\footnote{Note that this
  interpretation is not applicable to the corkscrew structure of the
  IRAS~16342 jet: the diameter of the corkscrew is $\approx
  0.25$~arcsec or 500~au at 2~kpc, which is incompatible with an
  orbital period of $\sim 50$~yr in a stellar binary system}.
Further examples of well-studied YSO jets driving high-speed H$_2$
outflows include HH212 (Correia et al. 2009), and the Class 0/I
outflows of HH34 and HH1 (Garcia Lopez et al. 2008). 
 
\subsubsection{Evidence for accretion discs}
There is considerable evidence that protostellar jets are intimately
linked with the presence of accretion discs, across a wide range of
source masses and evolutionary stages from Class 0 to Class III. 
The correlation between outflow
momentum and accretion rate (e.g. Richer et al. 2000 for Class 0
sources) has led to the concept of outflows as accretion-driven winds
from magnetized circumstellar discs (see reviews by K\"{o}nigl et
al. 2000 and Shu et al. 2000). It has been known for some time that the
momentum and energy associated with many pre-PN outflows is too high
to result from purely radiation-driven winds (Bujarrabal et al. 2001),
and MHD models involving the launching of jets from accretion discs
have been advanced in the evolved star case too (e.g. Soker \&
Rappaport 2000; Livio \& Soker 2001; Garc\'{i}a-Arredondo \& Frank
2004). In this case the formation of an accretion disc implies the
presence of a binary companion (see De Marco 2009 for a review of
binary interaction scenarios).

Using our extinction-corrected flux measurement of $I_{\rm 1-0
  S(1)}=2.97\times 10^{-16}$~W~m$^{-2}$, taking $d=2$~kpc,
and assuming that the $1-0$~S(1) line 
will contribute approximately 10
per cent of the total H$_{2}$ flux\footnote{see fig.~3 of Caratti o Garatti 
et al. (2006)} gives 
$L_{\rm H_2} = 0.37$~L$_{\odot}$. The H$_2$
luminosity of IRAS~16342 is therefore of the same order of magnitude as
the majority of jet-active Class 0 and I protostellar sources studied
by Caratti o Garatti et al.  (2006). In the protostellar sources
$L_{\rm H_2}/L_{\rm bol}\sim 0.04$ with the luminosity deriving from
the accretion process, so that $L_{\rm bol} \sim L_{\rm acc}$. If the
molecular outflow in IRAS~16342 is accretion powered then the
accretion luminosity, $L_{\rm acc}\sim 10$~L$_{\odot}$, is similar to
that of a Class 0 protostellar source.

Bright CO bandhead emission and permitted atomic lines such as
Na~{\sevensize{I}}, Ca~{\sevensize{I}}, Mg~{\sevensize{I}} have been
observed around embedded protostellar sources associated with HH jets
and are thought to be tracers of active accretion discs
(e.g. Antoniucci et al. 2008). In a VLT-SINFONI study of seven HH jet
systems associated with Class I sources, Davis et al. (2011) find that
six sources (HH 34-IRS, HH300-IRS, SVS13, HH 72-IRS, HH999-IRS, HH
26-IRS) show strong CO bandhead emission as well as the atomic
emission lines. In most cases the CO emission is coincident with the
jet source locations, strengthening the association with the central
accretion process, whereas the H$_2$ emission lines are offset from
the source and therefore appear to trace the collimated outflows. 
In IRAS~16342, the CO bandheads and Na~{\sevensize I}
line are seen in the scattered continuum so that their origin lies close to
the illuminating source and therefore the jet source region. This
provides further evidence that the jet in IRAS~16342 may be associated
with an active accretion disc, with the CO bandhead emission arising
either in the disc or in an accretion flow (Martin 1997). As the
mass-losing star in IRAS~16342 evolves to higher temperatures and 
lower mass-loss
rates, we may expect this molecular signature of accretion to dimisnish.
It is interesting to note that there is little sign of CO bandhead emission
in the spectrum of CRL~618 (Thronson 1981), which has a much hotter B0
central star, a H{\sevensize II} region, and seems to be at a more advanced 
pre-PN phase. 


\section{Conclusions}

Using integral field spectroscopy we have detected and imaged a fast, bipolar 
H$_2$ outflow from IRAS~16342-3814 (the Water Fountain) with the following 
properties:
\begin{enumerate}
\item We detect the $v=1-0$ S(0) to S(3), Q(1) to Q(4) and $v=2-1$
  S(1) and S(3) lines of H$_2$ in the $K$-band. Line diagnostics
  indicate shock-excitation with a gas temperature of 1400~K assuming
  LTE. There is some evidence for a higher temperature component of
  gas.
\item The H$_2$ emission peaks are located on the bipolar axis of the
  system at PA~68\degr~but offset further from the source than the
  continuum peaks. The emission in the NE lobe is red-shifted relative
  to systemic and that in the SW lobe blue shifted.
\item Accounting for an inclination of the bipolar axis to the plane
  of the sky of 45\degr, the deprojected axial velocity of the
  H$_2$-emitting gas (away from the source)  is 153 and $184\pm 24$~km~s$^{-1}$ 
in the NE and
  SW lobes, respectively.
\item We see asymmetric line profiles, with the red-shifted H$_2$
  emission in the NE lobe having a red wing and the blue-shifted
  emission in the SW lobe a blue wing. These wings become more
  pronounced with decreasing distance from the source, which we find
  difficult to account for in terms of a terminal bow shock model.
  Instead we suggest that the shocks may arise in fast-moving clumps,
  accelerated by the precessing jet. A clump interpretation may also
  provide a mechanism by which the H$_2$ can be excited in a low-velocity
  shock without dissociating, but exhibit the observed high-velocity
  bulk motion.
\item From the $1-0$~S(1) line we estimate that the mass of shocked
  H$_2$ is $M_{\rm H_2}=1.9\pm 0.5 \times 10^{-4}$~M$_{\odot}$ giving
  an outflow rate of $\dot{M}_{\rm H_2}=1.7\pm 0.7 \times
  10^{-6}$~M${_\odot}$~yr$^{-1}$, comparable with the mass
  flux estimated from CO observations.
\item The luminosity in H$_2$, $L_{\rm H_2}= 0.37$~L$_{\odot}$, is
  similar to that of Class~0 protostellar outflows which are thought
  to be powered by accretion discs. Evidence that the jet
  and outflow in IRAS~16342 are associated with accretion is provided by the
  detection of CO bandheads and a Na~{\sevensize I} feature in
  emission. These features are also observed in jet-active
  protostellar sources and are thought to trace hot gas
  in the inner regions of an accretion disc or flow.
\end{enumerate}

The similarities between the bipolar H$_2$ outflow of IRAS~16342-3814
and those of protostellar sources support the possibility that
molecular outflows at both ends of the main sequence may be powered by
accretion-driven jets. In this way evolved stars nearing the ends of
their lives may once again relive the energetic days of their youth.

\section*{Acknowledgments}
This work is based on observations with ESO telescopes at Paranal
Observatory under programme ID 075.D-0429(A).  We thank K. Lowe and
the staff of 
Paranal Observatory for assistance
with these observations. We thank Michael Smith for helpful
discussions on H$_2$ shock models.

\end{document}